# Low-Cost System for Automatic Recognition of Driving Pattern in Assessing Interurban Mobility using Geo-Information

Oscar Romero, Aika Silveira Miura, Lorena Parra and Jaime Lloret

**Abstract:** Mobility in urban and interurban areas, mainly by cars, is a day-to-day activity of many people. However, some of its main drawbacks are traffic jams and accidents. Newly made vehicles have pre-installed driving evaluation systems, which can prevent accidents. However, most cars on our roads do not have driver assessment systems. In this paper, we propose an approach for recognising driving styles and enabling drivers to reach safer and more efficient driving. The system consists of two physical sensors connected to a device node with a display and a speaker. An artificial neural network (ANN) is included in the node, which analyses the data from the sensors, and then recognises the driving style. When an abnormal driving pattern is detected, the speaker will play a warning message. The prototype was assembled and tested using an interurban road, in particular on a conventional road with three driving styles. The gathered data were used to train and validate the ANN. Results, in terms of accuracy, indicate that better accuracy is obtained when the velocity, position (latitude and longitude), time, and turning speed for the 3-axis are used, offering an average accuracy of 83%. If the classification is performed considering just two driving styles, normal and aggressive, then the accuracy reaches 92%. When the geo-information and time data are included, the main novelty of this paper, the classification accuracy is improved by 13%.

## 1. Introduction

In 2018, the World Health Organization reported 1.35 million yearly deaths globally caused by vehicle accidents [1]. The speed at which a vehicle travels directly influences the risk of a crash and the severity of injuries, as well as the likelihood of death resulting from that crash [2]. Therefore, real-time detection of abnormal driving patterns could effectively prevent such fatal accidents by alerting drivers to potential predicted dangerous scenarios. A driving pattern is how you drive a vehicle under different external factors and personal attributes. In the context of this paper, it is also defined as driving behaviour or driving style. In addition, real-time detection of abnormal driving patterns becomes a valuable approach for reporting the detected risk cases to a transportation management centre [3]. These actions include excessive speeding, improper following, erratic lane changing, and improper turns [4]. Three key contributing factors to road traffic accidents are human error, vehicle capabilities, and road infrastructures [5]. Therefore, it is essential to carry out studies regarding driving patterns in traffic. The decisions made during a trip depend on these conditions and are expressed in behavioural patterns or driving styles. Over the last three decades, several studies have aimed to classify driving styles using different tools such as self-reports or observed kinematic behaviours [6]. However, driving style is also analysed to suggest how to decrease the vehicle's fuel consumption [7], save battery in electric vehicles [8], improve driving health status [9], or for traffic surveys [10].

Several previous research projects have examined methods for classifying aggressive driving using different data and methods [4]. The main purpose of these studies is to identify behaviour patterns in order to recommend safer manoeuvres. Many companies offer expensive products on the market to determine the driver's driving pattern, mainly for the logistics and commercial sectors [11–13].

This paper presents a system for recommending changes in the driving pattern, taking into account the driving behaviour modelled by an ANN, using a low-cost sensor system with no installation requirements. The changes will include the vehicle speed, acceleration, and turn pattern. The system consists of a node device, physical sensors, a liquid-crystal display (LCD), a speaker, and a battery. Thus, the driver can be alerted in real time to increase or decrease the vehicle speed and acceleration patterns in order to maintain the safety standards defined by the model. The study includes driving tests for training and validation models carried out in the region of Valencia, Spain. Pattern recognition methods are used to identify the different types of driving styles. Three ways of driving related to speed, acceleration and turn patterns, geo-information (location of the car in terms of latitude and longitude), and time data were included. Since it has a local node execution system, there is no need to send data to the cloud or servers for automatic recognition of driving patterns, allowing offline operation. Nonetheless, in order to improve the ANN, the system allows the storage of information, and the possibility of sending it to the cloud once the driver has reached the destination. As far as we are concerned, no other paper has tested the use of geo-information combined with acceleration, speed, or turn data in driving style recognition. The main reason for including the geo-information is based on the expected variation of driving behaviour in different locations (for example, differences in driving patterns in low-lying areas versus ascending or descending slopes). In addition, with an extensive database, the geo-information can be used to identify the expected driving behaviour assigned to the type of road, even without roadmaps. This allows our system to operate offline without depending on the roadmap's external data.

The rest of the paper is structured as follows. Section 2 outlines the current solutions for assessing mobility on interurban roads. The proposal, including the mobility areas, the hardware, the driving tests, and the ANN establishment, is defined in Section 3. Section 4 describes the results of the driving tests analysing the variables. The impact, novelty, and progress beyond the state of the proposed system and its results are discussed in Section 5. Finally, Section 6 summarises the main conclusion of this paper and provides our future work.

## 2. Related Work

Currently, there are few low-cost monitoring solutions for drivers. However, studies propose low-cost alternatives using the existing sensors in cell phones to create applications [14], or coupled with other sensors such as Global Positioning System (GPS) and an accelerometer [5]. Accelerometer data provide insight into the longitudinal and lateral movement of the phone, while the onboard GPS receiver provides us the location data in terms of latitude and longitude [5]. In addition, with the advance in technology, some smartphone models have built-in muti-sensors, such as data collectors on the market. Since smartphones have increasingly become popular in the recent years and blended into our daily lives, more and more smartphone-based vehicular applications [15–17] have been developed in intelligent transportation systems [18]. Studies address driving event detection solutions to warn vehicle drivers about dangerous situations or manoeuvers while driving using sensors and smartphones [3,5,19–21] or neural networks for driving pattern detection [22,23].

Artificial Neural Networks (ANNs) are used to generate profiles of drivers manoeuvers and styles for the traffic simulation models. In this regard, ANN proves to be a powerful modelling technique as they allow the approximation of arbitrary nonlinear functions of complexity. Moreover, they allow the trade-off between fitting and interpolation depending on the number of neurons in the hidden layer [24]. The human driving models produce distributions over actions rather than maximum likelihood predictions, allowing stochastic predictions and the evaluation of statistical risk [25]. Therefore, there are different methodologies to infer and warn the driver in real time about unsafe manoeuvers that affect driving, dangerous turns, or high speeds [26].

The current accessibility of low-cost sensor systems and the importance of driver behaviour in the fields of vehicle energy saving, fuel economy, and driver safety brings out many studies using technologies in these fields. A portion of these studies is analysed in this section to identify the gap in the existing solutions.

Manzoni et al. used a model based on some variables such as acceleration and vehicle position to distinguish between aggressive and non-aggressive driving behaviour [27]. Another study described neural car-following models based on naturalistic driving data and outlined a general methodology for constructing such models. Moreover, some studies used facial features to aid in driver accident prediction. They combined both vehicle dynamics and driver face analysis for accident prediction [28].

Other research fields played an important role in developing studies of human driving behaviour. Álvarez et al. [8] presented a system for estimating an electric vehicle's battery consumption (and therefore the remaining charge) by considering the driving behaviour modelled through ANN, only using sensors typically included in current smartphones. In addition, the purpose of [29] was to better understand the nature of potential driving difficulties in individuals with a high-functioning autism spectrum disorder. Shino et al. [30] proposed indices that detect drivers' deviated states. The system considers drivers' judgment processes and uses the road environment and a naturalistic driving behaviour database. To achieve this objective, they focused on drivers' speed choice behaviour around curve situations, and they formulated the speed choice process [30].

Nevertheless, no other studies were found using low-cost external node sensor systems for driving, that send alerts, with no installation requirements. However, some others using smartphones are more common as a low-cost alternative. For example, Nguyen et al. [3] first proposed a novel approach: dynamic basic activity sequence (DAS) matching, a combination of machine learning and threshold-based methods for identifying normal and abnormal driving patterns. Secondly, they presented an efficient framework for recognising specific abnormal driving patterns, for instance, weaving, sudden braking, etc. [3]. In the study by Paleti et al. [19], the smartphone sensor data were used to compute microscopic traffic measures that were surrogated for these average driving patterns, and that were subsequently correlated with crashes.

To summarise, there are different approaches for pattern recognition in driving behaviour in the literature, which use artificial intelligence models and create internal driver alert systems. Nevertheless, it is still difficult to find prototypes using low-cost sensor systems. This paper addresses this gap in the literature by assembling a prototype that can record data collected by other coupled sensors, such as GPS and G-sensor, to warn the vehicle driver in real time about safe speed manoeuvres.

The current solutions aimed at identifying driving patterns are based on the use of devices connected to the car and sensing data about the engine. Those solutions have high cost and are specific to certain cars, with it not being possible to exchange the device from one car to another. The existing solutions based on those devices cannot be compared with our proposal regarding the economic cost. The potential benefits of the proposed system are the following:

• the low cost of our proposal benefits the users because of its lower cost;

• the high number of users generates a larger database providing more accurate decisionmaking rules by the ANN;

• the acquisition of the proposed system in areas with low incomes, generally rural areas, ensures that the generated database will cover rural areas and less populated places.

Other benefits not linked to the low price include the alerts in the form of audible signals when the driver changes the driving style, the continuous learning based on the ANN and the generated data, and the inclusion of geo-information and time data.

## 3. System Proposal

Although several systems have been provided to improve the driver's safety and efficiency, most rely on cameras or sensors in the cars. Nonetheless, a low-cost external system with no installation requirements is not found. Our system is based on a node that gathers information about a series of parameters which can be used to recognise the driving pattern of the driver in order to offer recommendations to enhance the safety of mobility. Artificial intelligence (AI) is used to identify the driver's driving style, specifically ANNs. In order to avoid unnecessary delays, dependence on coverage, and the extra cost of using internet service providers, the ANN will be executed locally in the node without relying on a cloud computing system.

*3.1. Mobility Areas Characterisation*

We have identified four different mobility areas to adjust our system (see Figure 1). Two of the identified areas belong to urban areas (area 1 and area 4), which are characterised by a lower speed limitation and a higher number of regulations affecting the driving behaviour, such as STOP signals, traffic jams, traffic lights, and pedestrian walkways. The main difference between both areas is that in residential areas, the roads are mostly composed of a single lane in each direction and overtaking other cars is not allowed. In workplaces, roads are generally characterised by more than one lane in each direction and surpassing other cars is allowed. Regarding interurban roads, two different areas are identified according to the type of road: conventional on-road (area 2) and highway (area 3). There are fewer aspects affecting the regulation on highways than the conventional road, which generally have two or more lanes. In those lanes, surpassing other cars is allowed. Meanwhile, when there are one or two lanes in conventional roads, surpassing other cars is generally allowed if the conditions are favourable. In some areas with a single lane, if conditions (visibility, curves, slopes, etc.) are not favourable, surpass other vehicles is not permitted. There are also differences in the most common vehicles using the roads in each area. The most crucial difference is the farming tractor and similar machinery using conventional roads. Other significant differences, such as lane capacity, traffic volume, road hierarchy, and surface quality, might vary considerably both within and between the aforementioned mobility areas. Nonetheless, these four mobility areas are considered for the purpose of this paper and to simplify the different mobility areas. The number and characteristics of the mobility areas can be increased in the future. A summary of the different characteristics of urban and interurban roads can be seen in Figure 1.

In order to train and verify the proposed system, a complex scenario of an interurban area, the conventional road, is considered. On the one hand, the variability of conditions affecting highway driver patterns are more limited. On the other hand, the urban areas might be too complex for the initial approximation to the proposed system. In the future, the system will need to be tested in more areas and include new areas and subareas with different characteristics.

*3.2. System Description*

The system's hardware is composed of a node device, a battery, two physical sensors, a liquid-crystal display (LCD), and a speaker. All the components are assembled in a single box which can be deployed on the car dashboard.

The physical sensors gather data about the driving pattern and its variation along the path. The physical sensors included in the system are a g-sensor (also known as an accelerometer or gyroscope) and a GPS sensor. Specifically, the selected sensors were the GY-521 MPU-6050 [31] (G-sensor) and the GPS GY-NEO6MV2 [32] (GPS-sensor). With the G-sensor, it is possible to obtain acceleration data for the 3-axes, defined in the paper as fax, fay, and faz, expressed in m/s2. The orientation of the coordinate system's axes concerning the vehicle is the following. The x-axis is aligned with the forward direction of the vehicle. The y-axis is tangential to the forward direction of the vehicle. Finally, the z-axis is aligned with the gravitational acceleration. In addition, the turning speed of rotation for the 3-axis axis can be calculated as fgx, fgy, and fgz, expressed in degrees/s. The G-sensor has a user-programmable accelerometer; the selected range was ±2 g. Thus, it offers a sensibility of 16,384 LSB/g. Concerning the GPS sensor, the data of the following parameters can be gathered: geo-information

(including latitude (◦, ",'"") and longitude (◦, ",' "), time (hh:mm:ss), and velocity (km/h). The sensors, with their respective connection in the protoboard, can be seen in Figure 2; in particular, the GPS sensor and antenna in Figure 2a, and G-sensor in Figure 2b.

The node is responsible for converting the signal of both sensors into the value for each one of the mentioned parameters. In addition, the node also powers the sensors. The node is configured to read data from the sensors every second. Sensed data are displayed on the LCD as a series of numbers. Moreover, all sensed data are stored in an SD card, and they can be uploaded to the cloud for their future use. In the node, the driving pattern recognitions are also performed using the sensors' data as input for an ANN. According to the result of the ANN, which can be a normal (Nor), conservative (Con), or aggressive (Agg) driving pattern, the node will trigger an acoustic advice for the driver. The data stored in the SD card can be included as tagged data in the ANN. Once the driver has finished driving, the data from the SD card can be downloaded to a computer and tagged by a cloud computing service which will adjust the ANN according to the last sensed data. In order to ensure that the selected node has enough computational capacity to execute the ANN, a Raspberry Pi 3 [33] was chosen (see Figure 3).

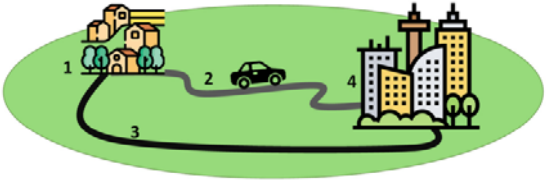

Figure 1. Identified areas and their characteristics affecting driver's patterns.

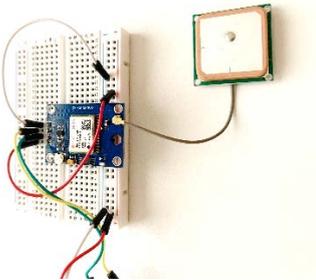
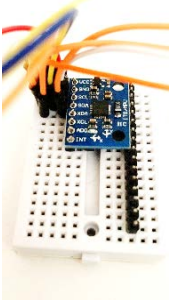

(a)          (b)

Figure 2. Sensors included in the system: Global Positioning System (GPS) sensor (a), and G-sensor (b).

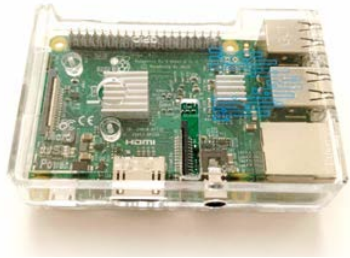

Figure 3. Selected node: Raspberry Pi 3.

As an overview, Figure 4 summarises the components and functions of the different elements of the system. The assembled prototype used in the test can be seen in Figure 5.

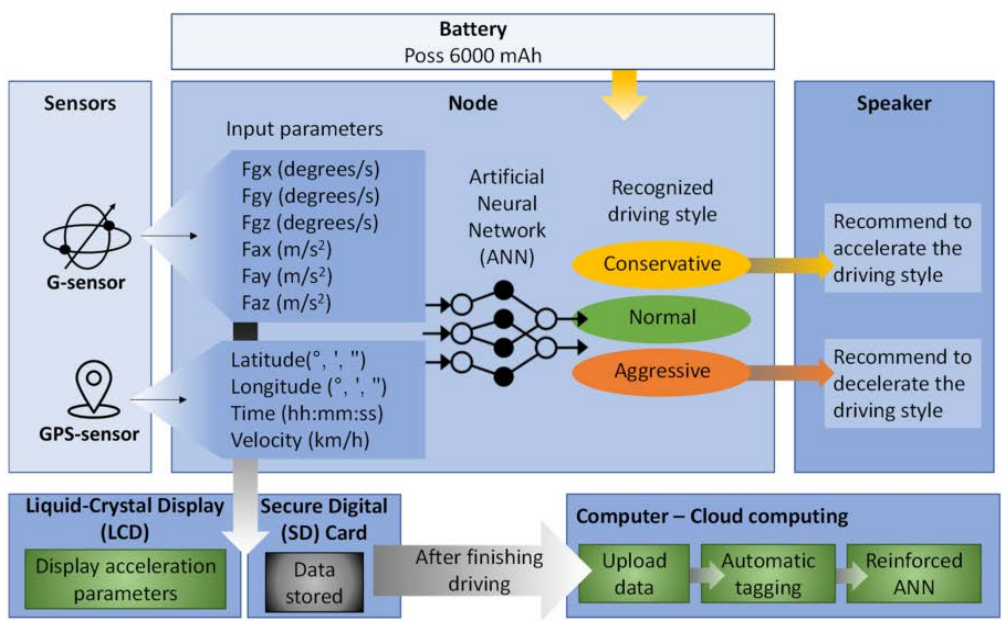

Figure 4. Summary or proposed system.

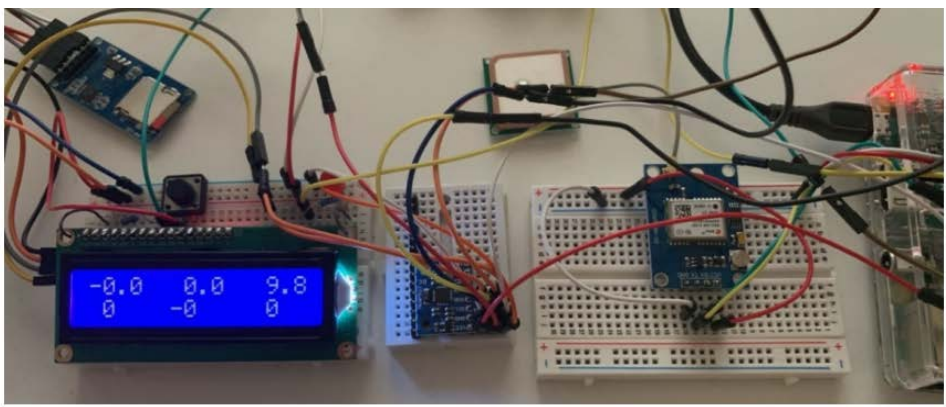

Figure 5. Assembled system.

*3.3. Driving Tests*

A series of driving tests have been conducted to generate data that can be used to test and validate our system. The tests were conducted in the region of Valencia (Spain); more specifically, two municipalities, Montserrat and Picassent. The area selected for the driving test includes 10 km of road connecting a residential area (in Monte Rosado from Montserrat town) and a workplace (in Picassent). The route has an average duration of 10 min and includes roundabouts, speed bumps, yields, STOPs, minor traffic jams, and slopes that affect driving.

In order to simulate different driving patterns, a driver drove the route with three different driving modes: Nor, Con, Agg. First, the driver drove the route in a regular mode, simulating the correct driving pattern, the Nor mode. Acceleration and braking were conducted typically, and the roundabouts were driven regularly. The velocity of the car was slightly below the limit of the road. The second driving mode, the Con driving pattern, includes slow and careful acceleration and braking, and roundabouts were driven exceptionally carefully. The velocity of the car was below the road limit. The last driving mode, simulating an Agg driving pattern, consists of abrupt acceleration, braking, and sharply driven roundabouts. The car velocity was equal to or slightly above the limit of the road.

In order to gather the variability of conditions, each route and driving pattern was repeated by the same driver three times. The driver had been asked to drive as similarly as possible in the three repetitions and to report differences in the routes regarding the traffic.

*3.4. ANN Establishment*

In order to find the combination of variables that more accurately recognises the driving pattern, the variables were divided into different groups. The velocity, geo-information, and time-related data were always included. Nonetheless, the data from the gyroscope (fax, fay, faz, fgx, fgy, and fgz) were added in lots. For the first test, all the data from the gyroscope were added. For subsequent tests, only fa or fg were included. In the first test, the ANN consisted of an input layer with ten neurons, two hidden layers, and an output layer with three neurons (the three driving patterns) (see Figure 6a). Meanwhile, for the second and third ANN, there were seven neurons in the input layer, two hidden layers, and an output layer with three neurons (see Figure 6b); while in the second ANN, the information of fa was added for the third ANN, and the added information belonged to the fg data.

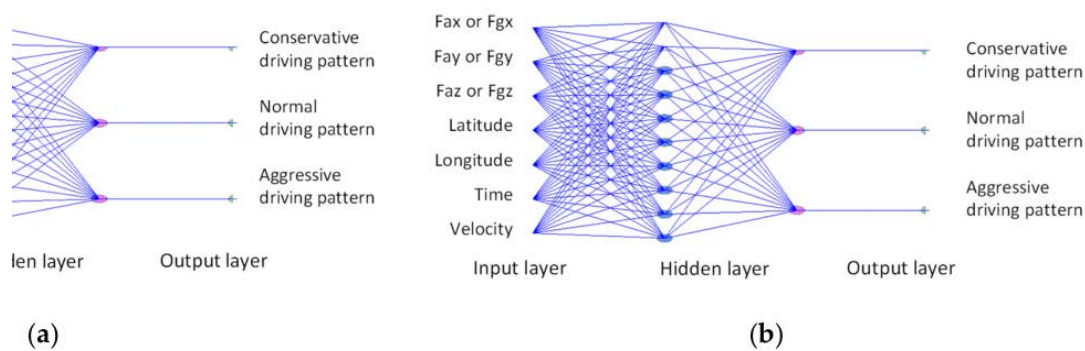

Figure 6. ANN design for the automatic recognition of driving parameters (a) using ten parameters and (b) using seven parameters.

## 4. Results

This section details our test results and the ANN's accuracy in tagging the driving pattern. The first step, described in the first subsection, verified whether there were differences in the variables analysed in the three driving styles. Then, the variables were included in the ANN to verify the accuracy of the proposed system, the prototype, and the ANN. The analysis, comparison, and discussion of the obtained results can be found in Section 5.

## 4.1. Variation of Studied Parameters

### 4.1.1. Velocity

The velocity registered every second was analysed in order to evaluate the differences among the driving patterns tested. Non-parametric statistics were used since the parameter does not follow a normal distribution. The Kruskal–Wallis test indicated that the velocity variance was different for the analysed driving patterns. The result of the Kruskal–Wallis test was a p-value equal to 0. The medians for each driving style were 59.3, 68.5, and 79.4 km/h for the Con, Nor, and Agg driving styles, respectively. The Bonferroni procedure pointed out that the differences between the three groups were statistically different, with 95% confidence. The geographic distribution of velocity along the route for the first repetition of the route for each one of the driving patterns can be seen in Figure 7. We can see in the figure the differences indicated by the statistical analysis. The colours indicate the value of the velocity at every point of the route.

### 4.1.2. Turning Speed

The turning speed data were analysed to find out if there were differences among the driving patterns tested. Since the $f_{gx}$, $f_{gy}$, and $f_{gz}$ do not follow a normal distribution, non-parametric statistics were used. Starting with $f_{gx}$, the Kruskal–Wallis test indicated that there was no statistically different variance among the analysed driving patterns. The result of the Kruskal–Wallis test was a p-value equal to 0.175679. For $f_{gy}$ and $f_{gz}$, the results indicated that no statistically significant differences were found for their variance. The p-values were equal to 0.115028 and 0.795761 for $f_{gy}$ and $f_{gz}$, respectively. The distribution of $f_{gx}$ along the route for the first repetition of the route for each one of the driving patterns can be seen in Figure 8. The colours indicate the value of the $f_{gx}$ at every point.

### 4.1.3. Acceleration

The gathered acceleration values were analysed to determine if there were differences among the driving patterns tested. Again, non-parametric statistics were used since the $f_{ax}$, $f_{ay}$, and $f_{az}$ do not follow a normal distribution. For $f_{ax}$, the Kruskal–Wallis test indicated that the variance of the $f_{ax}$ was not statistically different from the analysed driving patterns. The result of the Kruskal–Wallis test was a p-value equal to 0.59873. Regarding the $f_{ay}$, the results also pointed out no differences between driving patterns, with a p-value of 0.450834. Nonetheless, for the $f_{az}$, the Kruskal–Wallis test showed that its variance was different for the analysed driving patterns. The p-value was equal to 0.0135474, which confirms that the difference was statistically significant. The medians for each driving style were 9.84, 9.82, and 9.89 m/s2 for the Con, Nor, and Agg driving styles, respectively. According to the Bonferroni procedure, the differences between the three groups were statistically different with 95% confidence. The geographic distribution of $f_{ax}$ along the route for the first repetition of the route for each one of the driving patterns can be seen in Figure 9. The colours indicate the value of the $f_{ax}$ at every point.

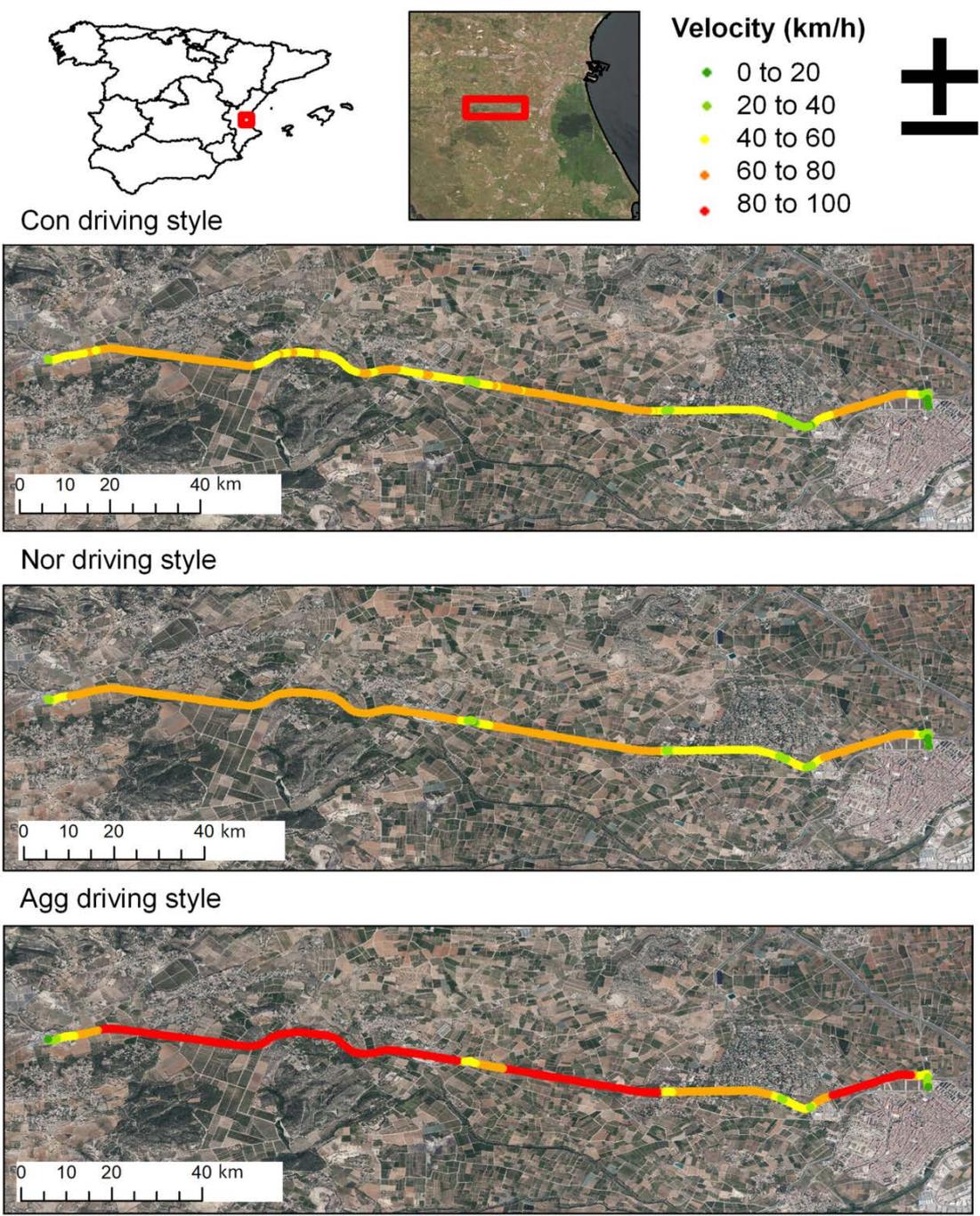

Figure 7. Variation of velocity along the route for the three driving patterns.

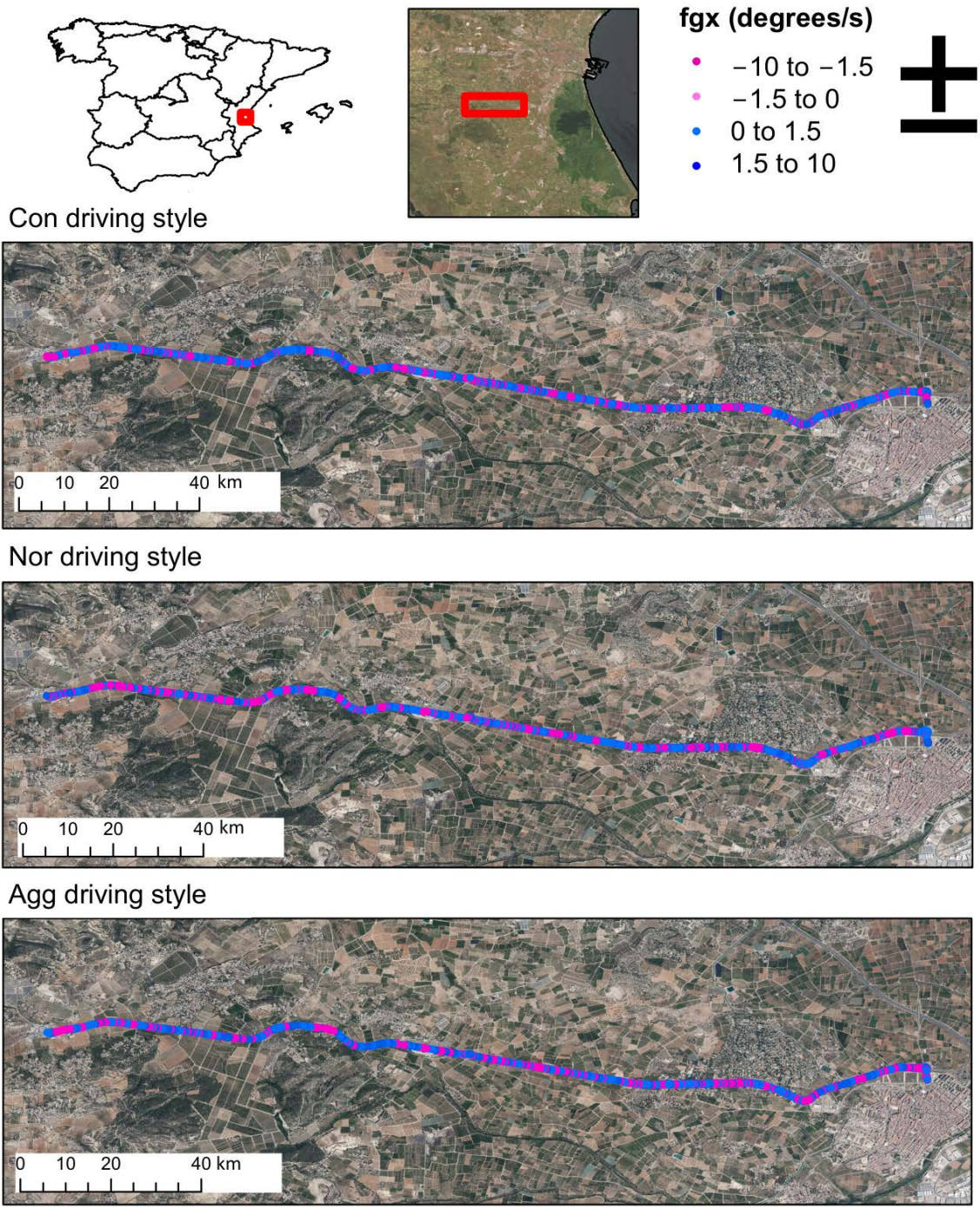

Figure 8. Variation of fgx along the route for the three driving patterns.

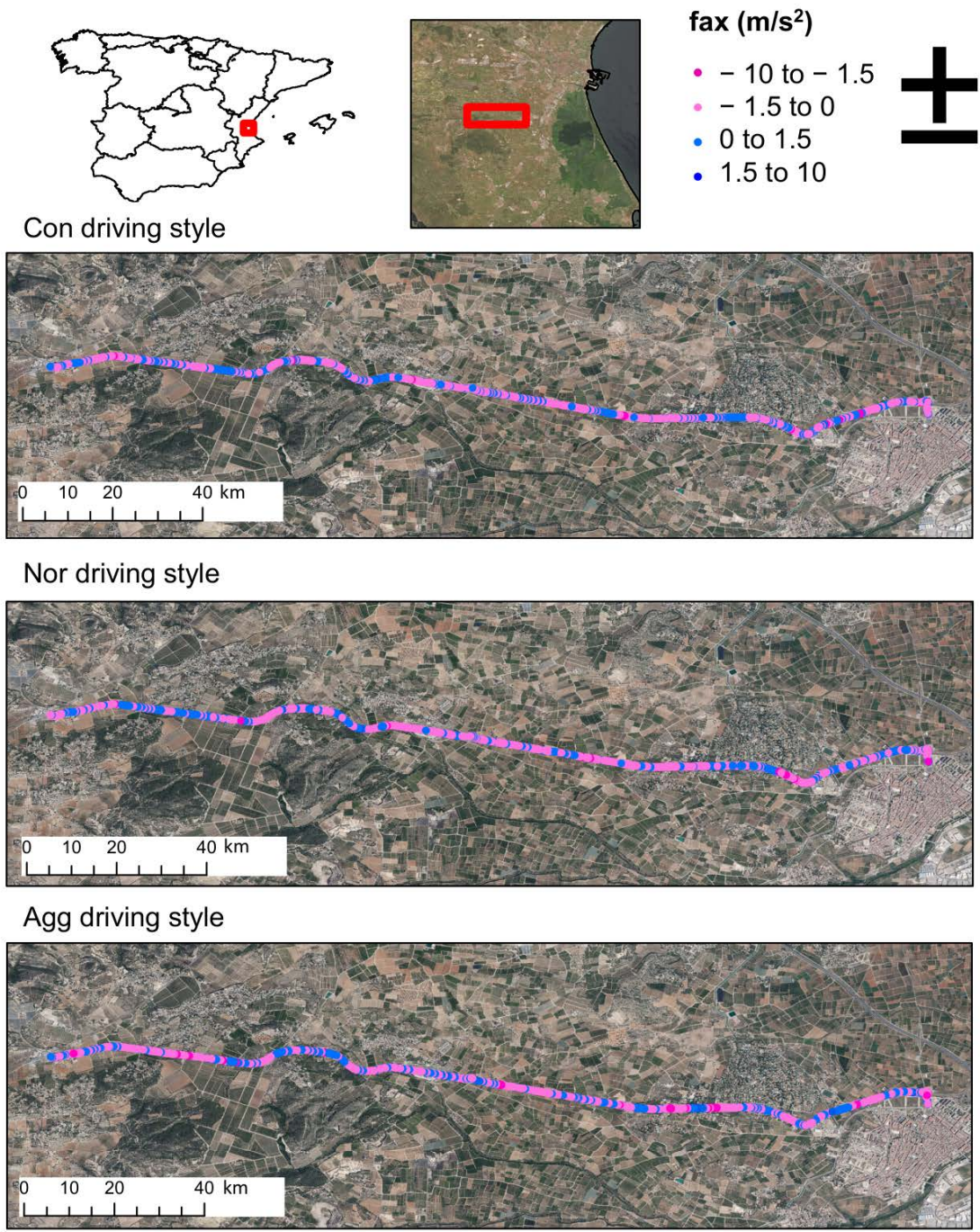

Figure 9. Variation of fax along the route for the three driving patterns.

## 4.2. ANN Performance

The data gathered in the test, a total of 5208 registers, were subdivided into two groups: the training groups to train the ANN and the validation group. Initially, the data of two first repetitions for each driving pattern were used to train the ANN and the data of the third test for its validation. Nonetheless, even though the driver has driven as similarly as possible in the repetitions, the differences between the routes were too significant. This means the ANN can accurately classify the dataset training, with accuracies above 90%, but most of the validation dataset was incorrectly classified, with accuracies below 29%. Thus, we have divided the data randomly into two datasets. Different sizes of training and validation datasets were selected (500, 1500, 2500, 3500, and 4500 registers for the training dataset).

Given the expected variability due to randomly divided data, each combination was tested ten times to have average training and validation accuracy values.

The results, in terms of classification accuracy for the different training and validation datasets, can be seen in Figure 10. The results indicate that the maximum accuracy, regardless of training dataset size, was attained when the input parameters for the ANN were: velocity, geo-information (latitude and longitude), time, fgx, fgy, and fgz. The results, including velocity, latitude, longitude, time, fax, fay, and faz, were similar but slightly inferior. Finally, the accuracy when all parameters are included was considerably inferior to the other cases. The maximum accuracies were achieved when the training dataset size was larger. The average accuracies for the validation dataset in these cases were 71.89, 82.12, and 83, 82% when including both fa and fg, only fa, and only fg. In order to allow the future comparison of the results, the confusion matrix obtained in training and validation datasets for an average case, which includes only fg, is shown in Tables 1 and 2. Table 1 depicts the results for a training dataset with an accuracy of 83.11%, while Table 2 presents the validation results with an accuracy of 82.49%.

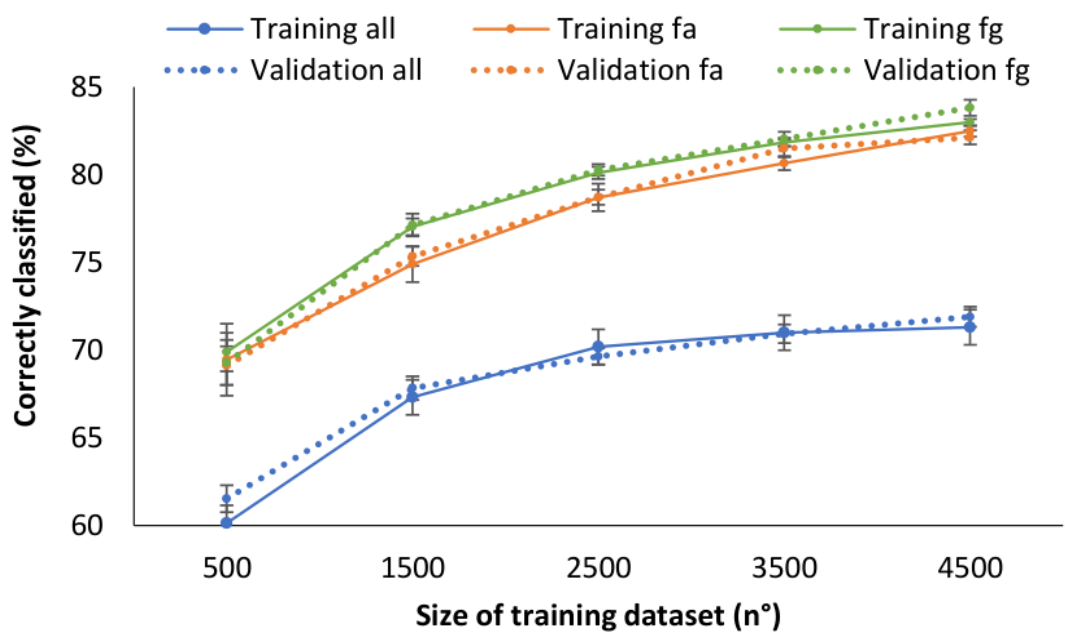

Figure 10. Summary of correctly classified data and the different sizes of training datasets when the different combinations of variables are used.

Table 1. Confusion matrix for training dataset when velocity, geo-information, time, fgx, fgy, and fgz are included.

| Type | Size | Classified as Con | Classified as Nor | Classified as Agg |
|---|---|---|---|---|
| Con | 1717 | 1516 (88.29%) | 165 (9.61%) | 36 (2.10%) |
| Nor | 1484 | 250 (16.85%) | 1145 (77.16%) | 89 (6.00%) |
| Agg | 1299 | 98 (7.54%) | 122 (9.39%) | 1079 (83.06%) |

Table 2. Confusion matrix for validation dataset when velocity, geo-information, time, fgx, fgy, and fgz are included.

| Type | Size | Classified as Con | Classified as Nor | Classified as Agg |
|---|---|---|---|---|
| Con | 278 | 255 (91.73%) | 15 (5.40%) | 8 (2.88%) |
| Nor | 234 | 42 (17.95%) | 177 (75.64%) | 15 (6.41%) |
| Agg | 196 | 20 (10.20%) | 24 (12.24%) | 152 (77.55%) |

# 5. Discussion

In this section, we will discuss the results comparing the differences between the accuracy of our system and the existing solution in the literature in the first subsection in a table. The second subsection deals with the potential benefits of the presented proposal on mobility. Finally, the limitations of the proposal, as well as the future functionalities, are detailed in the last subsection.

*5.1. Progress beyond the State of the Art*

In Section 2, we have already mentioned that there are a few proposals based on sensors capable of recognising the driver style. Only [27] presents a proposal in which the driver pattern was identified based on data with the same purpose as our system. Nonetheless, this proposal includes data about vehicles, such as throttle openings, as input. This is a significant difference from our systems since we have proposed a system that can be used in different vehicles and needs no physical connections with the vehicle. In addition, in [27], no real driving tests were performed; only a driving simulator was used. The accuracy of [27] is 86.6%, which is slightly higher than the accuracy reported in Table 2. Nonetheless, as in some papers, only two driving styles were analysed (aggressive and normal); we have adapted our datasets in order to allow comparison with only two classes. The first option is to delete the Con dataset, in which case we have an accuracy of 89.64%. The other option is to merge the Nor and the Con datasets; in this case, the accuracy reaches 92.43%. In both cases, we improve the results reported in [26].

Other examples of the combined use of smartphones and in-vehicle sensor data are presented in [34–37]. In all cases, real driving test data were used. In [36,37], only two driving patterns were analysed, obtaining maximum accuracies of 76.2 and 89.7% for [36,37]. Our results are very similar (when the Con dataset is not used) or improve (when the Con dataset is merged with the Nor dataset) accuracies even with no data from the in-vehicle sensors. Regarding the results of [35], three driving styles were analysed, and an accuracy of 92.16% was attained. Our accuracy for the three styles classification is slightly inferior. Nonetheless, our system has more flexibility than the one proposed in [35]. In [34], the authors achieve an accuracy of 99.99% with five different driving style patterns. Nonetheless, the classification method is much more complex than the proposed one and probably cannot be embedded in a Raspberry Pi 3. In addition, their results are based on data which required specific equipment connected to the car, which restricts the use of the prototype, requiring the addition of specific sensors according to the car.

We found a few examples in which only smartphone data were used [37,38]. Real driving tests were conducted in both papers. In [37], many variables sensed by smartphone sensors were included, and an accuracy of 66.7% is achieved when data were classified into two driving style patterns using machine learning. Our results improve the accuracy reported in [37]. Finally, ref. [38] includes data from an accelerometer (fax, fay, and faz) and GPS, and two driving styles are tested. Several techniques for data classification were compared, and the best accuracy was obtained with the support vector machine (SVM). The accuracy was 94% for the normal and 90% for the dangerous datasets. No data on overall accuracy is given. The accuracy for our datasets is 95%, and 85% for Con + Nor and Agg datasets. Thus, we get better accuracy for the Nor dataset, but lower accuracy for the Agg dataset. Considering that SVM is more complex than ANN, and that authors of [38] have also used a Radial basis function network with 20 hidden neurons and a radius of 5 in their test, with accuracies of 87% and 88% with dangerous and normal driving patterns, our proposal has better overall accuracy. All the data about current solutions are summarised in Table 3.

Table 3. Comparison of obtained results with existing solutions.

| Included Parameters | Classification Method | Driving Tests | Diving Patterns | Max. Accuracy | Ref. |
|---|---|---|---|---|---|
| [1] Vehicle speed and throttle opening | SVM and S3VM | Simulator (closed circuit) | Aggressive and normal | 86.6% | [27] |
| [1] Acceleration, gravity, revolutions per minute (RPM), speed, and throttle of the vehicle | CNN with three convolutional | Real tests (no information on the type of road) | Normal, aggressive, distracted, drowsy, and drunk | 99.99% | [34] |
| [1] Vehicle speed, acceleration, and throttle opening | Fuzzy c-means (FCM) clustering and SVM | Real tests, including urban and interurban road | Aggressive, conservative, and normal | 92.16% | [35] |
| [1] Vehicle speed, engine speed, engine load, throttle position, acceleration, gyroscope data, magnetic field | SVM and ANN | Real tests hilly extra-urban, main extra-urban, and urban roads | Normal and unsafe | 89.7% | [36] |
| GPS (location, speed, bearing), 3D acceleration, orientation, compass, gyroscope, linear acceleration, gravity, rotation vector, illumination, and air pressure | Machine learning technique | Real tests (no information on the type of road) | Relatively aggressive and relatively calm | 66.7% | [37] |
| [1] Previous plus engine coolant temperature, engine load, engine RPM, throttle position, and speed | | | | 76.2% | |
| Accelerometer (fax, fay, and faz) and GPS | Several best results with SVM | Real tests in an urban area | Dangerous and normal | 90% and 94% | [38] |
| Velocity, geo-information, time, fgx, fgy, and fgz | ANN | Real test in conventional road | Con, Nor, Agg<br>Nor and Agg<br>Nor (Nor + Con), Agg | 83.11%<br>89.64%<br>92.43% | Our proposal |
| Velocity, fgx, fgy, and fgz | ANN | Real test in conventional road | Con, Nor, Agg<br>Nor and Agg<br>Nor (Nor + Con), Agg | 69.10%<br>74.20%<br>79.06% | |

[1] The use of data from in-vehicle sensors removes the possibility of the use of these proposals in other cars.

The main novelty of our proposal is the use of geo-information and time data as a variable. In [38], GPS data were used only to re-calibrate the vector speed. In our proposal, these data can be used to reinforce the system. In scenarios with no information about the location of STOPs, curves, speed bumps, or slopes, the system will learn that in these areas (latitude and longitude values), it is normal to decrease the speed. In addition, the inclusion of time as a parameter will allow the system to identify daily or weekly patterns, such as dense traffic to access the working areas in the morning, reduced velocity near schools in the morning, or dense traffic to leisure areas at weekends, etc. These data endow our proposal with learning capabilities which are entirely new and necessary for its use in real and changing scenarios. In fact, we can identify that our system is already learning the effect of position (geo-information) and time on driving behaviour. If those variables are removed, accuracies decrease by 13%, and by 9% if time is not included.

Several papers used smartphone data to detect driving events using similar parameters, such as [4,5,38–40]. Nonetheless, the objective of our paper is deeper and more complex. Other articles present the use of similar devices to estimate fuel consumption, such as [41]. Those examples show the potential future uses of the proposed prototype. In those cases, the geodata are only used to correct the velocity [37,38], or no information about their use is given [40,41]. Only one case [5] has used the geo-information as input variables to determine driving events. Nevertheless, no information on time data is used.

In summary, the analyses of existing solutions and their accuracies confirm the novelty of our proposal, including geo-information and time data. In addition, they allow us to confirm that the obtained accuracies are aligned with the existing literature when similar classification methods are used.

*5.2. Impact of the Presented Proposal on Mobility*

Based on the novelty of our proposal and the reasonable accurate classification of driving styles, the impact that this prototype might have is clear. On the one hand, the prototype can be used in personally owned vehicles without the need of replacing devices when the system is changed from one vehicle to another. In this case, the benefit of the proposed system is to help the driver to keep in the Nor driving style, minimising the risks of accidents and maximising the efficiency of combustible, or energy, use.

On the other hand, the system can be used in professional vehicles. In this case, several possible uses can be identified. First of all, the potential use of this system in training vehicles will help the learners keep a Nor pattern in accelerating, turning, and approximating STOPs, curves, speed bumps, or slopes.

Secondly, it can be used in transport vehicles which will provide a rating of the professional drivers, helping the managers to provide their workers with accurate recommendations or training courses to maximise their rating.

Finally, considering the low cost of the system, its simplicity, and the fact that it does not require a connection with the car, the proposed system can be quickly adopted by drivers. As the number of drivers increases, the datasets uploaded to the cloud to train the ANN will be enlarged, creating additional rules which can be added to the nodes, increasing the system's accuracy.

*5.3. Limitations of the Current Proposal*

The current proposal, including the prototype and the ANN, was tested in real environments, particularly an Audi A1 car. During the trials, a battery (Power bank Poss 6000 mAh) was employed to power the node. The battery consumption observed during the trials was less than 30% of the initial energy after more than three hours of driving. Estimations indicate a possible lifetime of 10 h. Although the ANN was not running during the test, we can confirm that the prototype is fully operative, and the battery consumption assures that it can be used in cars even for long trips. ANN was not running during the test because there were no tagged data. The tests served to generate data, which is tagged later. Nevertheless, more powerful batteries, with more than 20,000 mAh, can be found on the market nowadays. A second possibility is to use the vehicle to power the node, allowing its operation regardless of travel time.

Nevertheless, the main limitation of our proposal is that the prototype and the ANN were only used in a single car. Thus, we need to extend the dataset not only to other cars (such as micros, sports utility vehicles, hatchbacks, pickups, and coupes, among others), but also to other vehicles (such as vans, trucks, and motorcycles). Since we expect that the type of vehicle will significantly impact the fa and fg values, new datasets will be necessary to train a new ANN. Two possibilities arise at this point. The first possibility is to join the datasets of different vehicles, including the type of vehicle as a new variable in a new ANN, which can be used in all vehicles interchangeably with a simple indication of the type of the current vehicle. The second possibility is to have different datasets and train several ANN for every kind of vehicle. In that case, the type of vehicle will not be a variable but will be required to apply the rules of the ANN of this type of vehicle. The major limitation of the second option is that the commercial prototypes will require a larger storage capacity in order to store the different ANNs, and probably the required size of datasets will be larger.

For future scenarios, the requirements of the system to get an ANN which can learn in real time from other devices located in other vehicles are:

• Ensure connectivity with enough bandwidth and low delay to exchange data with other connected devices and/or with a database in the cloud;

• Power the device with the battery of the car since the higher data rate transfer will require a high power consumption;

• Generate a method to tag the data in real time from the vehicles in order to update the ANN rules.

Different methods to tag the data are envisaged. (1) The first option is to define two default operation modes for the system. The first operation mode will be based on identifying the driving pattern. Meanwhile, the second available operation mode will be based on generating tagged data based on previously identified driving patterns. (2) The second method is based on correcting and tagging abnormal data (Con and Agg data) by including a microphone and a series of voice commands. Those commands will rectify the automatically tagged data when the driver realises that the ANN does not correctly tag the driving pattern. Thus, the result of data classification of the system with the voice commands can be assumed as tagged data, which can be input data for other running devices.

## 6. Conclusions

Mobility in urban or interurban areas is still causing many fatalities on the road. Even though some new vehicles are endowed with systems capable of sensing and recognising their environment to reduce the risk of accidents, several cars do not have those technologies yet. Existing systems are mostly based on in-vehicle sensors installed from the factory. The related literature shows that most car manufacturers solve this problem by adding a diagnosis module. Nonetheless, this solution cannot be easily standardised since modules are specific for every manufacturer and model.

This paper proposes a system independent of the car manufacturer and model, which relies only on data gathered from the crafted prototype. The system can identify the driver style based on the ANN model running on a node in an offline mode. The main novelty of our system compared with existing ones, in addition to its flexibility, is the use of geoinformation and time data to learn the differences in driving due to the heterogeneity of the road (STOPs, curves, speed bumps, or slopes). The overall accuracy of our system with all the variables and two diving styles is 92.43%, which decreases by 9% if time data are removed, and by 13% if time data and geo-information data are not included.

Future work will be linked to increasing the datasets by including urban scenarios and new vehicles. Moreover, it will also be combined with the use of distributed databases in order to compare the gathered data [42]. In addition, the use of SVM as a classification method will be tested. The fact of increasing the number of sensors of our devices in order to sense environmental data, which can affect driver behaviour, will be studied. Finally, we will analyse the potential benefit of combining sensed data with real-time information from Google Maps (or from other vehicles [43]) to consider if the system will benefit from having connectivity to propose an enhanced version.